\documentclass{PoS}
\usepackage[utf8]{inputenc}
\usepackage{verbatim}
\usepackage{epsfig}
\title{QCD Thermodynamics with Domain Wall Fermions}

\ShortTitle{QCD Thermodynamics with Domain Wall Fermions}

\author{\speaker{Michael Cheng} for the RBC and HotQCD Collaborations\\
        Columbia University\\
        E-mail: \email{michaelc@phys.columbia.edu}}

\abstract{
We present our recent studies of the pseudo-critical temperature, $T_c$, of 
QCD using domain wall fermions.  Domain wall fermions have the advantage that
they preserve exact $SU(2)$ chiral symmetry at finite lattice spacing in
the limit that $L_s \rightarrow \infty$.  The RBC Collaboration has
performed a set of dynamical calculations at $L_s = 32$ and $N_t = 8$ using 
the Iwasaki gauge action with two light quarks ($m_l a = 0.003$) and one 
strange quark ($m_s a = 0.037$).  A clear signal for the 
crossover transition can be seen in the light chiral susceptibility, as well 
as in the Wilson line.  However, at $L_s = 32$, the residual chiral symmetry 
breaking is not yet fully under control.  
We also present preliminary results from the HotQCD Collaboration with 
$N_t = 8$ and $L_s = 96$, where the effects of the residual chiral symmetry
breaking are reduced compared to $L_s = 32$.
         }

\FullConference{The XXVI International Symposium on Lattice Field Theory\\
		 July 14-19 2008\\
		 Williamsburg, Virginia, USA}

\begin{document}

\section{Introduction}

The location and nature of the QCD phase transition has been extensively 
studied using lattice techniques with various different fermion actions 
\cite{Cheng:2006qk,Bernard:2004je,Aoki:2006br,Chen:2000zu}.  
Recently, the most detailed studies of the transition temperature have been 
performed with different variants of the staggered fermion action
\cite{Cheng:2006qk,Bernard:2004je,Aoki:2006br}, which do not preserve the
full chiral symmetry of QCD at finite lattice spacing.  
Domain Wall Fermions (DWF)
on the other hand, allow the realization of exact chiral symmetry on the lattice,
at the cost of introducing of an auxiliary fifth dimension.

A study of the transition temperature with domain wall fermions 
was done for $N_t = 4, 6$ \cite{Chen:2000zu}.  However, it was found
that at such coarse lattice spacings, the DWF formulation begins to break
down, with unphysical effects that prevent the extraction
of a reliable estimate of $T_c$.

In this work, we present a study by the RBC Collaboration of the pseudo-critical
temperature, $T_c$, with domain wall fermions with $N_t = 8$ and fifth-dimensional
extent $L_s = 32$.  It is hoped that, at the finer lattice spacings needed for $N_t = 8$,
the large lattice artifacts that appear at $N_t = 4, 6$ are under better control.  We also
present preliminary results from the HotQCD Collaboration with domain wall fermions at
$N_t = 8$ with $L_s = 96$.

\vspace{-0.4cm}
\section{Simulation Details at $L_s = 32$}

For our study we utilize the standard DWF action with an Iwasaki gauge action.
The behavior of DWF at zero temperature has been extensively studied for 
this combination of actions; see refs. \cite{Antonio:2007tr, Allton:2008pn} 
for more details.

The Rational Hybrid Monte Carlo (RHMC) algorithm 
\cite{Clark:2004cp,Clark:2006fx}, an exact algorithm that satisfies
detailed balance and reversibility, is used to generate the gauge configurations.
A three-level nested Omelyan integrator
is used in the molecular dynamics evolution, with $\lambda = 0.22$.
The length of the molecular dynamics trajectories between Metropolis steps is
chosen to be $\tau = 1$.  The step size is tuned to achieve an acceptance rate of 
approximately $75\%$.  A spatial volume of $16^3$ is used for the finite temperature ensembles 
with $N_t = 8$.  For each value of the gauge coupling, we use a fixed value for the bare
light and strange quark masses ($m_l a = 0.003$ and $m_s = 0.037$).

1200 molecular dynamics trajectories were also generated
at $\beta = 2.025$ with a volume of $16^3 \times 32$ and $L_s = 32$, 
at the same quark masses used for the finite temperature 
ensembles.  These configurations were used to calculate the static quark potential, as
well as the meson spectrum.

From the static quark potential, we obtain a value for the Sommer parameter, $r_0/a = 3.08(9)$.  
Using $r_0 = 0.469(7)$ fm., this indicates a lattice scale of $a^{-1} \approx 1.3$ GeV at
$\beta = 2.025$.  The meson spectrum measurements, give a pion mass $m_\pi \approx 310$ MeV, while
the kaon mass is within 10\% of the physical value.

\vspace{-0.4cm}
\section{Finite Temperature Observables}
The observables that we use to probe the chiral properties at a given temperature
are the light and strange quark chiral condensates 
($\left<\bar \psi \psi_l\right>,\left<\bar \psi \psi_s\right>$), and
the disconnected part of the chiral susceptibility ($\chi_l, \chi_s$).
They are defined as:
\begin{eqnarray}
\left<\bar \psi \psi_q\right> & = &\frac{\partial \ln Z}{\partial m_q} = \frac{1}{N_s^3 N_t}\left<Tr(M_q^{-1})\right>\\
\chi_q & = & \left<(\bar \psi \psi_q)^2\right> - \left<\bar \psi \psi_q\right>^2
\end{eqnarray}
On all of our finite temperature configurations, we measure both the light
and strange chiral condensates every fifth trajectory, using 5
random sources per configuration to estimate $Tr(M_q^{-1})$.

\begin{figure}[t]
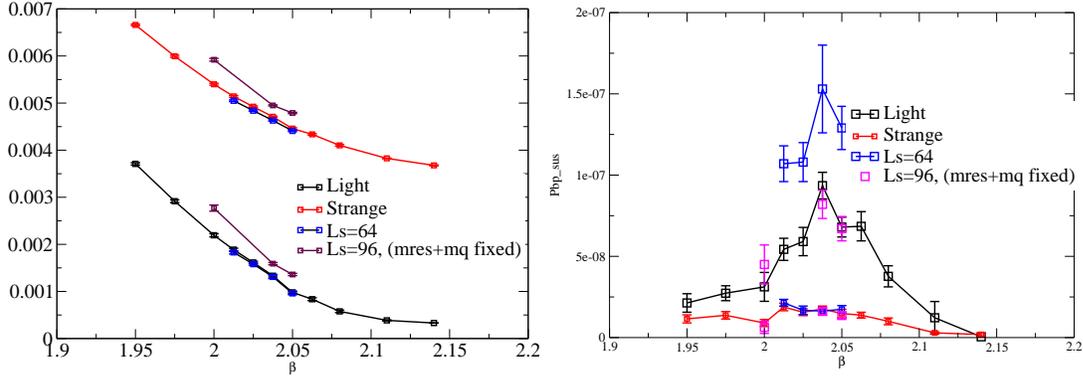

\begin{minipage}[c]{0.47\textwidth}
\includegraphics[width=\textwidth]{figs/pbp.eps}
\end{minipage}
\begin{minipage}[c]{0.47\textwidth}
\includegraphics[width=\textwidth]{figs/pbp_sus.eps}
\end{minipage}
\caption{On the left, $\left<\bar{\psi}\psi\right>$ for $L_s = 32, 64, 96$.  On the right, the disconnected
chiral susceptibility for $L_s = 32, 64, 96$.}
\label{fig:pbp}
\vspace{-0.5cm}
\end{figure}

Figure \ref{fig:pbp} shows the chiral condensate and
the disconnected part of the chiral susceptibility, respectively.  Examining 
the light and strange chiral condensate, it is difficult to precisely locate an
inflection point, which is the signal for a thermal crossover.
However, we can use the disconnected chiral susceptibility, a measure
of the fluctuations in the chiral condensate.  As seen in figure
\ref{fig:pbp}, there is a clear peak in the light disconnected susceptibility.
The results for the chiral condensates and the associated susceptibilities are
summarized in table \ref{tab:ls32_results}.  Fits to the peak region indicate
that $\beta_c = 2.041(5)$.

We also measure the observables that probe confinement, i.e. the Wilson line and its
associated susceptibility.  These are defined as:
\begin{equation}
\left<W\right>  =  \frac{1}{N_s^3}\sum_{\mathbf{x}}Tr \left(\prod_{t=0}^{N_t-1} U_{\mathbf{x}, t}\right);~~ \chi_W  =  \left<W^2\right> - \left<W\right>^2.
\end{equation}

The results for the Wilson line and Wilson line susceptibility are also given in table \ref{tab:ls32_results}.

\begin{table}[b]
\begin{tabular}{r@{.}l|c|cc|cc|cc}
\hline
\hline
\multicolumn{2}{c|}{$\beta$} & Traj. & $\left<\bar{\psi}\psi_l\right> ~(10^{-3})$ & $\chi_l ~(10^{-8})$ & $\left<\bar{\psi}\psi_s\right> ~(10^{-3})$ & $\chi_s ~(10^{-8})$ & $\left<W\right> ~ (10^{-3})$ & $\chi_W ~ (10^{-4})$\\
\hline
1&95   & 745  & 3.71(3) & 2.13(57) & 6.66(2) & 1.15(25) & 4.40(62) & 1.15(10)\\
1&975  & 1100 & 2.92(3) & 2.73(46) & 5.99(2) & 1.37(23) & 5.44(42) & 1.42(10)\\
2&00   & 1275 & 2.19(3) & 3.12(89) & 5.40(1) & 0.90(22) & 6.52(47) & 1.32(10)\\
2&0125 & 2150 & 1.89(3) & 5.43(68) & 5.14(2) & 1.88(23) & 9.02(53) & 1.46(5)\\
2&025  & 2210 & 1.62(3) & 5.91(87) & 4.92(2) & 1.56(21) & 10.18(61)& 1.45(8)\\
2&0375 & 2690 & 1.33(3) & 9.35(82) & 4.71(2) & 1.76(18) & 13.61(55)& 1.43(6)\\
2&05   & 3015 & 0.98(3) & 6.80(61) & 4.46(2) & 1.48(26) & 16.77(71)& 1.56(7)\\
2&0625 & 2105 & 0.84(3) & 6.85(90) & 4.34(2) & 1.38(18) & 18.22(86)& 1.72(9)\\
2&08   & 1655 & 0.58(3) & 3.77(65) & 4.10(3) & 1.00(21) & 25.91(129)&1.78(11)\\
\end{tabular}
\caption{Summary of finite temperature observables, as well as the number of trajectories generated.}
\label{tab:ls32_results}
\end{table}

\vspace{-0.4cm}
\section{Residual Mass}
One of the primary drawbacks of the current calculation is the rather
large residual chiral symmetry breaking for the parameters that we employ.
This manifests itself in a value for the residual mass, $m_{res}$
which is larger than the input light quark mass, $m_l a = 0.003$ over
almost the entire range of parameters in our calculation.

We have measured $m_{res}$ at $\beta = 2.025$ on the $16^3 \times 32$
zero temperature ensemble, which gives $m_{res} = .00665(8)$.  We have also
measured the residual mass on the finite temperature lattices.  These measurements
agree well with measurements on zero temperature lattices at nearby $\beta$.

\begin{figure}[t]
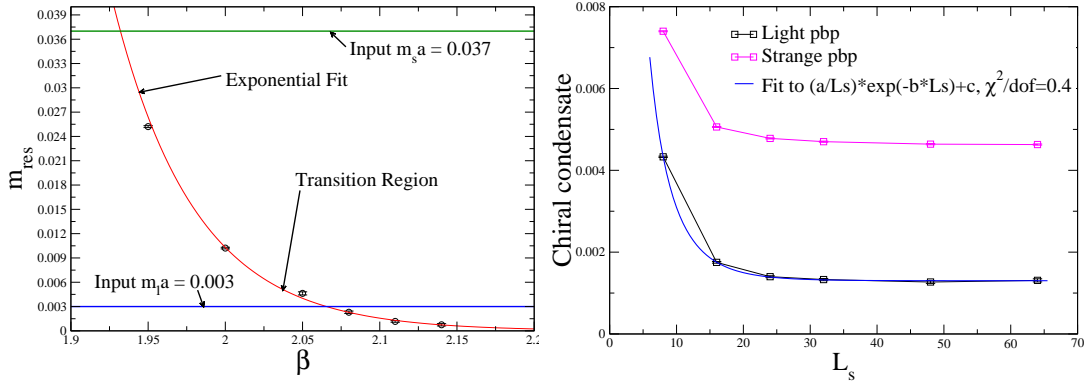

\begin{minipage}[c]{0.47\textwidth}
\includegraphics[width=\textwidth]{figs/mres.eps}
\end{minipage}
\begin{minipage}[c]{0.47\textwidth}
\includegraphics[width=\textwidth]{figs/pbp_b2.0375.eps}
\end{minipage}
\caption{On the left, the dependence of $m_{res}$ on $\beta$.  On the right, the dependence of 
$\left<\bar{\psi}\psi\right>$ on $L_s$ at fixed input quark mass $m_q a = 0.003$ at $\beta = 2.0375$.}
\label{fig:mres}
\vspace{-0.5cm}
\end{figure}

Figure \ref{fig:mres} shows how $m_{res}$ varies with $\beta$.
As we can see, $m_{res}$ has a strong, exponential dependence on $\beta$.
While we have chosen the input light quark mass $m_l = 0.003$ to be fixed
at the different $\beta$, the exponential dependence
of $m_{res}$ means that the effective light quark mass, 
$m_q = m_l + m_{res}$ changes significantly in the crossover region,
from $m_q \approx .0075$ at $\beta = 2.05$ up to $m_q \approx .013$ at
$\beta = 2.00$.

\vspace{-0.4cm}
\section{Chiral observables at varying $L_s$}
The shifting of the quark mass with $\beta$ results in a distortion 
of the shape of the susceptibility
curves that we use to locate the crossover transition.  In order to
understand how this varying mass affects our results, we have measured
the partially quenched chiral condensate with different $L_s$ and $m_l$ 
at various $\beta$.  In particular, we choose
$L_s = 64$ with the same input light quark, $m_l = 0.003$, while for
$L_s = 96$, we vary the input quark mass so that the total effective
quark mass $m_q$ approximately matches that for $L_s = 32$ at the chosen value 
of $\beta$.  For one value of the gauge
coupling ($\beta = 2.0375$), we measure $\bar{\psi}\psi$ with many
choices of valence $L_s$ and fixed input quark masses
$(m_l,m_s) = (0.003, 0.037)$.
Table \ref{tab:pbpLs} give the results of these measurements.
Figure \ref{fig:pbp} shows the results with
valence $L_s = 64$ and $L_s = 96$ in context with the $L_s = 32$ results.

\begin{table}[hbt]
\centering
\begin{tabular}{cc|ccc|ccc}
\hline
\hline
$L_s$ & $\beta$ & $m_l$ & $\left<\bar{\psi}\psi_l\right> ~(10^{-3})$ & $\chi_l ~(10^{-8})$ & $m_s$ & $\left<\bar{\psi}\psi_s\right> ~(10^{-3})$ & $\chi_s ~(10^{-8})$\\
\hline
8 & 2.0375 ~& ~0.003 & 4.33(2) & 2.39(22) & ~0.037 & 7.40(1) & 1.43(15)\\
16 & & & 1.75(2) & 4.05(40) & & 5.05(1) & 1.46(16)\\
24 & & & 1.40(2) & 5.89(63) & & 4.78(1) & 1.51(18)\\
48 & & & 1.28(3) & 11.4(14) & & 4.65(1) & 1.58(20)\\
\hline
64 & 2.0125 ~& ~0.003 & 1.83(4) & 10.7(9) & ~0.037 & 5.05(2) & 2.15(21)\\
& 2.025  & & 1.58(3) & 10.8(11) & & 4.84(2) & 1.67(25)\\
& 2.0375 & & 1.31(4) & 15.3(25) & & 4.63(1) & 1.63(20)\\
& 2.05   & & 0.96(3) & 12.9(14) & & 4.41(2) & 1.74(25)\\
\hline
96 & 2.025 ~& ~0.0078 & 2.03(2) & 5.59(92) & ~0.0418 & 5.29(1) & 1.27(20)\\
& 2.0375 ~& ~0.0063 & 1.59(2) & 8.22(75) & ~0.0403 & 4.95(1) & 1.61(19)\\
& 2.05 ~& ~0.0070 & 1.36(3) & 6.70(62) & ~0.0410 & 4.79(2) & 1.35(20)\\
\hline
\end{tabular}
\caption{Partially quenched measurements of $\left<\bar{\psi}\psi\right>$ at 
different $L_s$, $\beta$, $m_l$.}
\label{tab:pbpLs}
\end{table}

From Figure \ref{fig:pbp}, we see that holding the input quark mass fixed
at $m_l = 0.003$, while reducing $m_{res}$ by setting $L_s = 64$ does not have a 
significant effect on the chiral condensate.  On the other hand, with a larger input quark
mass and $L_s = 96$, the chiral condensate shifts appreciably.  Figure
\ref{fig:mres} shows the dependence of $\left<\bar{\psi}\psi\right>$
on $L_s$ at $\beta = 2.0375$.  For small values of $L_s$, there is a strong
dependence, but the chiral condensate quickly plateaus to an approximately
constant value for $L_s = 32, 64$, even though $m_{res}$ is still changing
significantly in this region.

In contrast to the chiral condensate, the disconnected part of the chiral
susceptibility depends on the total effective quark mass, 
$m_q = m_l + m_{res}$.
As seen in figure \ref{fig:pbp}, when the total quark mass is changed at $L_s = 64$,
the chiral susceptibility differs significantly from the measurements at $L_s = 32$.
However, when we keep the total quark mass $m_l + m_{res}$ fixed at $L_s = 96$,
the resulting chiral susceptibility agrees with $L_s = 32$.
Thus, while the chiral condensate is sensitive to the relative contributions from the
input quark mass and the residual mass, the chiral susceptibility is a function of
only the total quark mass, $m_q = m_l + m_{res}$.

\vspace{-0.4cm}
\section{Determining $T_c$ at $L_s = 32$}
\label{sec:Tc}
While the peak in the light chiral susceptibility is well-determined to be
$\beta_c = 2.041(5)$, there are several issues that need to be adressed in extracting
a physical value for the pseudo-critical temperature, $T_c$.

Since $m_{res}$ changes so drastically as a function of $\beta$, the chiral
susceptibility curve is distorted by the changing light quark mass.  Taking a
simple ansatz $\chi_l \sim 1/m_q$, we can adjust our data so that the bare quark
mass is fixed as $\beta$ varies.  This adjustment shifts $\beta_c$ to stronger
coupling, $\beta_c = 2.031(5)$.

We have determined the lattice scale at $\beta = 2.025$, which differs slightly from
 $\beta_c = 2.031(5)$.  Using a simple interpolation between our result at $\beta = 2.025$
and the results at weaker coupling\cite{Li:2006gra}, we obtain $r_0/a = 3.12(13)$ at
$\beta = \beta_c$.  The results in ref. \cite{Li:2006gra} also indicate that chiral extrapolation and
finite-volume effects add 4\% to this value, giving $r_0/a = 3.25(18)$, where the error bar has
been artificially inflated to include this 4\% in the uncertainty.

Since this calculation is done only at $N_t = 8$ and at one set of quark masses, we cannot
perform either the chiral or continuum extrapolation needed to obtain a value for $T_c$ at physical
quark masses in the continuum.
From ref. \cite{Cheng:2006qk}, we can estimate the effect of the chiral extrapolation to the
physical quark masses to be
about 5\%.  Ref. \cite{Cheng:2006qk} also found that the effect of the continuum extrapolation
is approximately 5\%, although it is obtained using the p4 action.
For our purposes, we estimate the error from the lack of continuum and chiral extrapolations 
to be 10\%.

Taking these errors into account, we obtain a value of $T_c r_0 = .406(23)(41)$, or $T_c = 171(10)(17)$ MeV,
where the first error bar takes into account all the systematic errors outlined above except for the chiral
and continuum extrapolation, which are reflected in the second error.

\vspace{-0.4cm}
\section{Results at $L_s = 96$}
\begin{table}[t]
\begin{tabular}{r@{.}l|c|cc|cc|cc}
\hline
\hline
\multicolumn{2}{c|}{$\beta$} & Trajectories & $m_l a$ & $m_s a$ & $\left<\bar{\psi}\psi_l\right> ~(10^{-3})$ & $\chi_l ~(10^{-8})$  & $\left<W\right> ~ (10^{-3})$ & $\chi_W ~ (10^{-4})$\\
\hline

1&9875 & 1395 & 0.00250 & 0.0407 & 2.15(3) & 8.7(14) & 6.1(5) & 1.34(9)\\
2&00   & 1485 & 0.00325 & 0.0415 & 1.97(3) & 8.2(12) & 5.4(6) & 1.29(9)\\
2&0125 & 1425 & 0.00395 & 0.0422 & 1.68(3) & 10.1(17) & 9.4(6) & 1.59(11)\\
2&025  & 1730 & 0.00435 & 0.0426 & 1.64(2) & 8.3(11) & 9.6(6) & 1.45(8)\\
2&0375 & 1630 & 0.00485 & 0.0431 & 1.33(3) & 9.5(10) & 13.6(6) & 1.45(9)\\
2&05   & 1565 & 0.00525 & 0.0435 & 1.21(2) & 6.7(10) & 14.4(6) & 1.43(9)\\
\end{tabular}
\caption{Finite temperature observables for $L_s = 96$}
\label{tab:ls96_results}
\end{table}
The primary drawback of the RBC Collaboration's calculation just described
is the rapidly changing residual mass as a function of $\beta$ in the
transition region.  As we have discussed, this means that the effective
quark masses in the strong coupling side of $\beta_c$ are significantly larger
than those on the weak coupling side.  This has the effect of distorting the
shape of the chiral susceptibility peak in the $L_s = 32$ calculation, making the
peak appear sharper and at weaker coupling than if we worked at fixed quark mass.

The HotQCD Collaboration's calculation seeks to address this flaw by working at
$L_s = 96$.  Utilizing $L_s = 96$ reduces $m_{res}$ approximately by a factor of 3.
It also allows us to choose the input quark masses ($m_l a$ and $m_s a$) so that
the total effective quark mass ($(m_l + m_{res}) a$ and $(m_s + m_{res}) a$) are
the same at each value of $\beta$ that is used.  For this calculation,
the light quark mass is chosen to be $1/6$ the strange quark mass at each value of
$\beta$, where the strange quark is chosen to be approximately physical.  The corresponding
bare quark masses are given in table \ref{tab:ls96_results}.  Otherwise, the lattice actions, 
the spatial volume, and the molecular dynamics algorithm used are all identical to
those used at $L_s = 32$.

\begin{figure}[b]
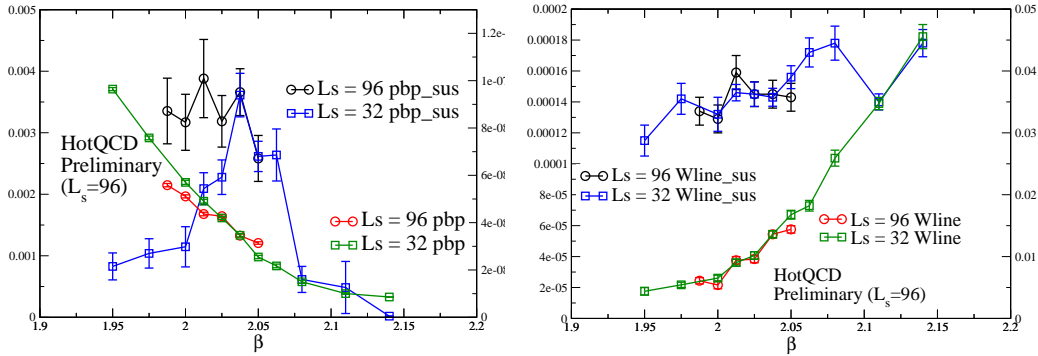

\begin{minipage}[c]{0.45\textwidth}
\includegraphics[width=\textwidth]{figs/dwfLs96_pbp_compLs32.eps}
\end{minipage}
\begin{minipage}[c]{0.45\textwidth}
\includegraphics[width=\textwidth]{figs/dwfLs96_wline_compLs32.eps}
\end{minipage}
\caption{On the left, a comparison between $L_s = 32$ and $L_s = 96$ of the light quark chiral condensate and chiral
susceptibility.  On the right, the same comparison for the Wilson line and its susceptibility.}
\label{fig:ls96_result}
\vspace{-0.3cm}
\end{figure}

Table \ref{tab:ls96_results} gives preliminary results for the chiral condensates, wilson line, and their
susceptibilities.  These results are also presented in figure \ref{fig:ls96_result}.  As expected, working
at fixed quark mass results in a chiral susceptibility where the peak is broader and
much harder to resolve.  In addition, there are some indications that $\beta_c$ is
at stronger coupling at $L_s = 96$ compared to $L_s = 32$, as expected, but the
statistical error makes this far from a certain conclusion.

\vspace{-0.4cm}
\section{Conclusions and Outlook}
\label{sec:conclusion}
We have presented two studies of the critical region of finite temperature QCD
using domain wall fermions at $N_t = 8$.  
The calculation at $L_s = 32$ by the RBC uses observables related
to chiral symmetry (i.e. chiral condensate and disconnected chiral 
susceptibility), to calculate the crossover transition temperature, giving
the result $T_c r_0 = .406(23)(41)$. or $T_c = 171(10)(17)$ MeV.  

The second calculation by the HotQCD calculation improves upon the RBC
calculation by using $L_s = 96$ to further reduce the residual chiral symmetry breaking,
while tuning the input quark masses so that the total bare quark mass (including $m_{res}$)
is fixed at each different value of $\beta$.  However, preliminary results
show that the peak is less sharply resolved compared to $L_s = 32$, so no reliable
estimate for $\beta_c$ can yet be obtained.

Data at a few additional $\beta$ at both stronger and weaker coupling are needed
to better resolve the shoulders of the chiral susceptibility peak (if
indeed such a peak exists).  In addition, a zero temperature calculation to determine
the lattice scale needs to be done in order to obtain a physical value for $T_c$ in MeV.

\vspace{-0.2cm}
\bibliography{lattice2008}

\section*{Acknowledgments}
The work presented here by the RBC Collaboration used computational resources provided by
the U.S. Department of Energy, Brookhaven National Lab, Columbia University, and RIKEN.  
The work presented by the HotQCD Collaboration was made possible by computational
resources provided by the Lawrence Livermore National Lab and the NNSA.  Both projects utilized
resources at the New York Center for Computational Sciences at 
Stony Brook University/Brookhaven National Lab, which is supported by the U.S. DOE under Contract 
No. DE-AC02-98CH10886 and by the state of New York.  This work was also supported in part by U.S. DOE grand number
DE-FG02-92ER40699.

\end{document}